\documentclass[12pt]{article}
\usepackage{graphicx}

\begin{document}

\title{On the Origin of Elementary Particle Masses}
\author{Johan Hansson\footnote{c.johan.hansson@ltu.se} \\
 \textit{Department of Physics} \\ \textit{Lule\aa University of Technology}
 \\ \textit{SE-971 87 Lule\aa, Sweden}}

\date{}

\maketitle

\begin{abstract}
The oldest enigma in fundamental particle physics is: Where do the observed masses of elementary particles come from?
Inspired by observation of the empirical particle mass spectrum we propose that the masses of elementary
particles arise solely due to the self-interaction of the fields
 associated with a particle. We thus assume that the
mass is proportional to the strength of the interaction
 of the field with itself. A simple application
 of this idea to the fermions is seen to yield a mass for the
neutrino in line with constraints from direct experimental
upper limits and correct order of magnitude predictions of mass separations
between neutrinos, charged leptons and quarks.
The neutrino interacts only through the weak force, hence becomes light.
The electron interacts also via electromagnetism and accordingly becomes
heavier. The quarks also have strong interactions and become heavy.
The photon is the only fundamental particle to remain massless, as it is chargeless. Gluons gain mass comparable to quarks,
 or slightly larger due to a somewhat larger color charge.
 Including particles outside the standard model proper, gravitons are not exactly massless, but very light due to their very weak self-interaction.
 Some immediate and physically interesting consequences arise: i) Gluons have an effective range $\sim 1$fm, physically explaining why QCD has finite reach
 ii) Gravity has an effective range $\sim 100$ Mpc coinciding with the largest known structures; the cosmic voids
 iii) Gravitational waves undergo dispersion even in vacuum, and have all five polarizations (not just the two of $m=0$), which might explain why they have not yet been detected.
\end{abstract}
The standard model of particle physics \cite{Glashow}, \cite{Weinberg}, \cite{Salam}, \cite{Gell-Mann}
is presently our most fundamental \textit{tested} \cite{Cahn} description of nature. Within the standard model
there are some 18 parameters (several more if neutrinos are non-massless) which cannot be predicted but must be supplied
by experimental data in a global best-fit fashion. There are coupling constants, mixing parameters, and, above all, values for the different
fundamental particle masses. The theory is silent on where and how these parameters arise, and even more speculative theories, such as string theory,
has so far not been able to predict (postdict) their values.
Even if the Higgs particle is confirmed, and the Higgs mechanism \cite{Higgs} is validated in one form or another, it still does not explain ``the origin of mass" as often erroneously stated. Unknown/incalculable parameters for particle masses are in the Higgs model replaced by equally unknown/incalculable coupling constants to the Higgs field; the higher the coupling the larger the mass, while no coupling to the Higgs field gives massless particles like the photon and gluons. So nothing is gained in the \textit{fundamental} understanding of masses. Fifteen of the free parameters in the standard model are due to the Higgs. Thirteen of them are in the fermion sector, and the Higgs interactions with the fermions are not gauge invariant so their strengths are arbitrary. So to make progress we must understand masses.

There is no hope of predicting elementary masses from renormalized quantum field theory as the very process of renormalization itself forever hides any physical mass-generating mechanism; the renormalized masses are taken as the experimentally measured values, \textit{i.e.} any possible physical connection for predicting particle masses is lost. But surely, nature herself is not singular, the infinities appearing in quantum field theory instead arising from the less-than-perfect formulation of the theory. If a truly non-perturbative description of nature would be found it might be possible to calculate particle masses from first principles, but we still seem far from such a description.

In this article we will instead take a more phenomenological approach, but still be able to deduce a number of physical results and some interesting consequences.

From standard (perturbative) quantum field theory the lowest order contribution to the self-mass is (see Fig.1)
\begin{equation}
\Delta m  = \alpha \int \bar{u} \gamma_{\mu} K(1,2) \gamma^{\mu} u e^{ipx_{12}} \delta (s_{12}^2) d^4 x,
\end{equation}
where the loop integral is logarithmically UV divergent $\propto log(\frac{1}{r})$ as the cut-off radius $r \rightarrow 0$.\footnote{Also for a classical electron of radius $r$, $\Delta m = C \alpha \propto \alpha$, but there the coefficient is linearly divergent $C \propto 1/r$. Additionally, the classical result is exact, \textit{i.e.} non-perturbative.} So (in perturbation theory) the contribution is divergent but as all gauge fields diverge in the same way, the \textit{quotients} are finite. (Another way would be to assume that there exists a ``shortest length" in nature that would serve as a natural cut-off and give finite integrals.) As an aside, as all expressions are relativistically invariant the usual relativistic factor $\gamma = 1/\sqrt{1 - v^2/c^2}$ is automatic if $v \neq 0$, \textit{i.e.} if we are not in the rest frame of the particle.
\begin{figure}
  \centering
  \begin{center}
\includegraphics {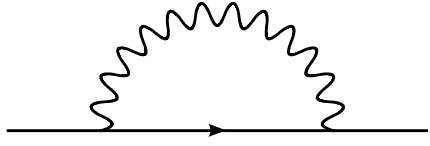}
\end{center}
  \caption{Feynman diagram for self-mass contribution from a gauge field (squiggly line). Each vertex contributes one charge factor $\sqrt{\alpha} \propto q$.}
\end{figure}

We will thus imagine the following pragmatic scenario; a quantum field without any charges corresponds to a massless particle; when charges, $q$, are attached the mass is $m \propto q^2 \propto \alpha$, where $\alpha$ is the relevant coupling constant. This significantly reduces the number of \textit{ad hoc} parameters. Also, the lagrangian can still be completely massless (as in the Higgs scenario), preserving attractive features such as gauge invariance that would be broken by explicit mass terms, the generation of mass being a secondary \textit{physical} phenomenon.

So we get
\begin{equation}
m(electron) \propto \alpha_{QED}
\end{equation}

\begin{equation}
m(quark) \propto \alpha_{QCD}
\end{equation}

\begin{equation}
m(neutrino) \propto \alpha_{QFD}
\end{equation}
where the dominating coupling constant is $\alpha_{QED}$ for quantum electrodynamics, $\alpha_{QCD}$ for quantum chromodynamics (strong interaction) and $\alpha_{QFD}$ for quantum flavordynamics (weak interaction).

If we now assume that all gauge fields give a contribution of roughly the same order of magnitude, so that the proportionality factors cancel up to a constant of order unity (coming from the different gauge groups), we get results for the quotients of elementary masses without having to know the exact (non-perturbative) contribution. Using the observed mass for the electron, and $\alpha_{QED} \sim 137^{-1}$, $\alpha_{QCD} \sim 1$, we get

\begin{equation}
m(quark) \simeq 50 \, MeV,
\end{equation}
(although physical quark masses are notoriously hard to define \cite{Hansson})
and pretending as if we knew nothing of the electroweak theory (in order not to get entangled with the Higgs mechanism again), using the old Fermi theory for weak interactions (or quantum flavordynamics, QFD) as appropriate for the low energies where observations of physical masses are actually made, using the \textit{physical} coupling derived from typical scattering cross sections or decay rates ($\tau^{-1} \propto \alpha^2$) , we get, using $\tau_{QFD}^{-1} \sim 10^{6} s^{-1}$ (\textit{e.g.} $\mu \rightarrow e \nu_{\mu} \bar{\nu_e}$) and $\tau_{QED}^{-1} \sim 10^{16} s^{-1}$ (\textit{e.g.} $\pi^0 \rightarrow \gamma \gamma$)

\begin{equation}
m(neutrino) \simeq 0.5 \times 10^{-5} MeV \simeq 5 \, eV.
\end{equation}
This is a \textit{prediction} resulting from our simple assumption, compatible with upper limits from direct experiments, whereas in the Higgs model \textit{no} predictions of masses are possible (being connected to free parameters).

We see that we immediately get the right hierarchy of masses, with the right magnitudes, which is encouraging considering the approximations made.

A clear indication of the relative effect of QED compared to QCD is seen in the case of pions;
$\pi^+$ and $\pi^-$ both have mass 139.6 MeV, while the neutral pion $\pi^0$ has a mass of 135 MeV.
The small difference $\Delta m = 4.6$ MeV, attributable to QED, predict a charge radius $\sim 1$ fm, consistent with scattering experiments using pions.

One issue still remaining is why not $m(Z) \sim m(neutrino)$ or $m(W) \sim m(electron)$. We take it as a sign that the intermediate vector bosons $W$ and $Z$ really are not fundamental, but instead are composite \cite{preons}, \cite{PreonTrinity}.

If we, disregarding renormalization issues, also include the graviton as the force carrier of gravity (which is expected to hold for weak gravitational fields) we see that QCD, QFD and gravity all should disappear exponentially at sufficiently large distances due to the non-zero physical masses of their force carrier particles, only electromagnetism (QED) having truly infinite reach as the physical mass of the photon is equal to zero, as the photon carries no charge. The range can be estimated by the Yukawa theory potential $e^{-\lambda m c/\hbar} /r$, giving $\lambda_{cutoff} \simeq \hbar/mc$. This gives for the gluon with bare mass zero (in the lagrangian), but physical mass $m(gluon) \neq 0$, the value $\lambda_{cutoff}(QCD) \simeq$ 0.3 fm, which explains why QCD is only active within nuclei, although the bare mass $m=0$ naively would give infinite reach as its coupling to the Higgs is zero. Despite what many thinks, this problem has not been solved \cite{Clay}.

For gravity the same calculation leads to $\lambda_{cutoff}(Gravity) \simeq 3 \times 10^8$ light years, or 100 Mpc, which happens to coincide with the largest known structures in the universe, the cosmic voids \cite{voids}. The corresponding graviton mass is $m(graviton) \simeq 5 \times 10^{-32}$ eV, well in line with the experimental upper limits \cite{nieto}. Another thing to keep in mind is that if/when gravity decouples, it will appear as if the universe accelerates when going from the coupled (decelerating) to the uncoupled (coasting) regime where distance $\geq \lambda_{cutoff}(Gravity)$, perhaps making dark energy superfluous as explanation for cosmic ``acceleration" \cite{Riess}, \cite{Perlmutter}. If masses really originate in this way it might be possible to include other interactions but the gravitational in an ``equivalence principle", hence perhaps opening the door to a unified description of all interactions.

$m(graviton) \neq 0$ has other peculiar effects: Gravitational waves of different wavelengths (energies) would travel at different velocities, smearing them out, the longer the wavelength the larger the effect. Also, not being strictly massless, gravitons (spin $s$=2) should have $2s + 1 = 5$ polarization states instead of the two conventionally assumed helicity states if massless. This might be why gravitational waves hitherto have escaped detection, as it would scramble their signature.

If we, just for the moment, tentatively reintroduce the perturbative running of coupling ``constants" (renormalization group) we obtain $m(graviton) \rightarrow \infty$ as $r \rightarrow 0$ implying that (quantum)gravity gets a dynamical cutoff for small separations, as an increasingly more massive quantum is harder to exchange, effectively making the interaction of gravity disappear in that limit, perhaps showing a way out of the ultraviolet divergencies of quantum gravity in a way reminiscent of how massive vector bosons cured the Fermi theory.

We have not addressed the known replication of particles into three generations of seemingly identical, but more massive, variants, the most exactly studied from an experimental standpoint being the three charged leptons, \textit{i.e.} ($e$, $\mu$, $\tau$), the electron and its heavier ``cousins" the muon and tauon.\footnote{Are there additional generations? Data on the decay width of the $Z$ indicate that there at least cannot be any additional light neutrinos. A fourth neutrino would have to be very massive $> m_Z /2 \simeq 45$ GeV. One might well ask if the generation structure is a true aspect of nature, or just a result of our incomplete understanding of the weak interaction \cite{PreonTrinity}.}

A straightforward way would be to introduce some ``generation charge" or quantum number, make \textit{e.g.} a power-law ansatz and fit to the observed values of the charged leptons and deduce the masses of neutrinos and quarks in the higher generations. That would, however, not bring us any closer to a true understanding.

A more promising way could be to assume that the stable elementary particles of the first generation are exact soliton solutions to the relevant quantum field theory, or its dual \cite{Manton}, whereas unstable higher generation elementary particles would be solitary wave (particle-like, but not stable) solutions to the said quantum field theory. Unfortunately, there are no known exact 3+1 dimensional soliton solutions to quantum field theories, with non-trivial soliton scattering \cite{Manton}. Another avenue would be to explore if Thom's ``catastrophe theory" \cite{Thom} (or other more general theories of bifurcation) applied to particle physics could spontaneously reproduce multiple generations, as it is known to include stable/unstable multiple solutions. Thom's theory states that all possible sudden jumps between the simplest attractors - points - are determined by the elementary catstrophes, and the equilibrium states of any dynamical system can in principle be described as attractors. As one attractor gives way for another the stability of the system may be preserved, but often it is not. It could be capable to generate masses spontaneously in a different and novel way compared to the Higgs mechanism. The different charges, \textit{i.e.} coupling constants, could define the control surface, whereas the actual physical mass would define the behavior surface. Sudden bifurcations could signify decay of previously stable elementary particles.

To summarize, our simple and physically compelling assumption that particle masses are solely due to self-interactions: i) Directly and simply gives the correct mass hierarchy between neutrinos, electrons and quarks. ii) Reduces the number of \textit{ad hoc} parameters in the standard model. iii) Qualitatively explains why the photon is the only massless fundamental particle, why QCD has short range, and why neutrinos are not strictly massless. iv) Gives testable predictions, \textit{e.g.} regarding gravitons (gravitational waves).

\end{document}